\documentstyle[12pt,epsf]{article}

\newcommand{\lsim}{\mathrel{\lower4pt\hbox{$\sim$}}
\hskip-12.5pt\raise1.6pt\hbox{$<$}\;}

\newcommand{\gsim}{\mathrel{\lower4pt\hbox{$\sim$}}
\hskip-12.5pt\raise1.6pt\hbox{$>$}\;}

\newcommand{\sss}{\scriptscriptstyle}

\newcommand{\be}{\begin{equation}}
\newcommand{\ee}{\end{equation}}
\newcommand{\bea}{\begin{eqnarray}}
\newcommand{\eea}{\end{eqnarray}}
\newcommand{\ba}{\begin{array}}
\newcommand{\ea}{\end{array}}

\newcommand{\pl}{Phys.\ Lett.}
\newcommand{\np}{Nucl.\ Phys.}
\newcommand{\prl}{Phys.\ Rev.\ Lett.}
\newcommand{\prd}{Phys.\ Rev.\ D}

\newcommand{\zp}{Zeit. \ Phys.} 
\newcommand{\ijmp}{Int.\ J.\ Mod.\ Phys.} 

\newcommand{\etal}{{\it et al}.,\ }

\begin{document}

\pagestyle{empty}
\hbox{\hspace{3.5truein} CEBAF-TH-96-01} 
\hbox{\hspace{3.5truein} BNL- }

\vspace{.7truein}
\begin{center}
{\large    \bf $R_b$ and $R_c$  in the Two Higgs Doublet Model}
{\large  \bf with Flavor Changing Neutral Currents} \bigskip

David Atwood,$^a$ Laura Reina,$^b$ and Amarjit Soni$^b$ \\
\end{center}
\medskip
\begin{flushleft}
$^a$Theory Group, CEBAF, Newport News, VA\ \ 23606 \\ 
$^b$Physics Department, Brookhaven National Laboratory, Upton, NY\ \
11973 
\end{flushleft}

\vspace{1truein}

\begin{quote}
{\bf Abstract}: A study of $R_b$ and $R_c$ is presented in the context
of a Two Higgs Doublet Model (2HDM) with flavor changing scalar
currents (FCSC). Implications of the model for the $\rho$-parameter
and for $b\to s\gamma$ are also considered. The experimental data on
$R_b$ places stringent constraints on the model parameters. The
configuration of the model needed to account for $R_b$ is found to be
irreconcilable with constraints from $b\to s\gamma$ and $\rho$. In
particular, if $R^{\rm exp}_b>R^{\sss{\rm SM}}_b$ persists then this
version of 2HDM will be ruled out or require significant
modifications. Noting that aspects of the experimental analysis for
$R_b$ and $R_c$ may be of some concern, we also disregard $R^{\rm
exp}_b$ and $R^{\rm exp}_c$ and give predictions for these using
constraints from $b\to s\gamma$ and $\rho$ parameter only. We
emphasize the theoretical and experimental advantages of the
observable $R_{b+c}\equiv \Gamma(Z\to b\bar b\mbox{ or } c\bar
c)/\Gamma(Z\to\mbox{hadrons})$. We also stress the role of
$R_\ell\equiv \Gamma(Z\to\mbox{hadrons})/\Gamma(Z\to \ell^+\ell^-)$ in
testing the Standard Model (SM) despite its dependence on QCD
corrections. Noting that in models with FCNC the amplitude for $Z\to
c\bar c$ receives a contribution which grows with $m^2_t$, the
importance and uniqueness of precision $Z\to c\bar c$ measurements for
constraining flavor changing $t\bar c$ currents is underscored.
\end{quote} 
\vskip 1truein

\newpage
\pagestyle{plain}
\section{Introduction and Summary}
\label{intro}

For the past several years precision studies at LEP have been
providing important confirmation to various aspects of the Standard
Model (SM) \cite{eps}. A notable exception that has emerged is the
decay of $Z\to b\bar b$. It has long been recognized that the $Zb\bar
b$ vertex is very sensitive to effects of virtual, heavy particles
\cite{pich}. Consequently, a deviation from the prediction of the SM
could prove to be a significant clue to {\it new physics}. It is,
therefore, clearly important to study $Z\to b\bar b$ in extensions of
the SM \cite{wells} and pursue the resulting implications. In this
paper we study these decays in a class of Two-Higgs-Doublet Models
(2HDM), called Model III \cite{sher}-\cite{mumutc}, which present a
natural mechanism for flavor changing scalar currents (FCSC).

Our focus is the branching ratio of $Z\to b\bar b$, i.e.\ \cite{eps}

\be
R_b \equiv \frac{\Gamma(Z\to b\bar b)}{\Gamma(Z\to\hbox{hadrons})}
\label{rbone} 
\ee

\noindent It is worth noting that, since $R_b$ is a ratio between two
hadronic rates, most of the electroweak (EW) oblique and QCD
corrections cancel between numerator and denominator, making it a
uniquely clean and sensitive test of the SM\null. Experiment finds
\cite{eps}:

\be
R^{\rm exp}_b = .2205\pm.0016   \label{rbexp}
\ee

\noindent whereas the SM prediction is \cite{eps}

\be
R^{\sss{\rm SM}}_b = .2156   \label{rbsm}
\ee

\noindent The difference, of about 3$\sigma$, is a possible indication
of new physics. We note, in passing, that the related decay $Z\to
c\bar c$ has also been measured albeit with appreciably less precision
\cite{eps}

\be 
R^{\rm exp}_c = .1543\pm.0074  \label{rcexp}
\ee

\noindent The SM prediction, on the other hand, is \cite{eps}

\be
R^{\sss{\rm SM}}_c =  .1724 \label{rcsm}
\ee

\noindent Thus $R^{\rm exp}_c$ also appears not to be consistent with
the SM although the deviation is milder (about $2.3\sigma$). It is
interesting to note that whereas $R^{\rm exp}_b$ is larger than
$R^{\sss{\rm SM}}_b$, $R^{\rm exp}_c$ is less than the SM
expectation. Note also that $R^{\rm exp}_b$ quoted above is obtained
by holding $R_c$ fixed to its SM value \cite{eps}.

Our findings are that if we take $R^{\rm exp}_b$ at its face value
then, while Model III can accommodate $R^{\rm exp}_b$, the model
parameters get severely constrained. In particular, the resulting
configuration of the model cannot be reconciled with the constraints
from the $\rho$-parameter and $Br(B\to X_s\gamma)$.

Several aspects of the $R_b$, $R_c$ experimental analysis are, though,
of concern. The results given above in eqs. (\ref{rbexp}) and
(\ref{rcexp}), include systematic errors and emerge from combining the
numbers from the four LEP detectors \cite{eps}. Since some of the
assumptions are common, treatment of the systematics can be
problematic. Also the errors for $R_b$ and $R_c$ are correlated
\cite{eps}. Indeed $R^{\rm exp}_b + R^{\rm exp}_c$ is consistent with
the SM accentuating the possibility that part of the effect may well
be resulting from misidentification of the flavors. In addition, the
observable $R_\ell$,

\be
R_\ell \equiv \frac{\Gamma(Z\to \mbox{hadrons})}{\Gamma(Z\to
\ell^+\ell^-)} \label{rell}
\ee

\noindent which is measured much more precisely than $R_b$ or $R_c$
and can be predicted in the SM using $\alpha_s(M_Z)$ deduced from
other methods (e.g.\ lattice and/or event shapes in $e^+e^-$
annihilation), is found not to be inconsistent with the SM, at
present.

In light of these reservations we also fix the parameter space by
using only the $\rho$-parameter and $Br(B\to X_s\gamma)$ and predict
$R_b$, $R_c$ and $R_{b+c}$ in Model III\null. In particular, in this
model, with constraints from the $\rho$-parameter and $Br(B\to
X_s\gamma)$, we find that $R_b$ cannot exceed $R^{\sss{\rm
SM}}_b$. Thus, if the current trend in the experimental numbers (i.e.\
$R^{\rm exp}_b > R^{\sss{\rm SM}}_b$) persists, this class of 2HDM
will be either entirely ruled out or require a significant alteration.

In passing we also emphasize the advantages of the observable
$R_{b+c}$

\be
R_{b+c} = \frac{\Gamma(Z\to b\bar b\mbox{ or } c\bar c)}{\Gamma
(Z\to \mbox{hadrons})} \label{rbplusc}
\ee

\noindent and give the predictions from Model III for $R_{b+c}$.

Finally, we stress the importance of precision determinations of $Z\to
c\bar c$ (i.e.\ $R_c$). In type III models its amplitude receives a
contribution which grows with $m^2_t$. A precise determination of
$Z\to c\bar c$, thus, constitutes a uniquely clean method for
constraining the flavor-changing $tc$ vertex that is of crucial
theoretical concern.

\section{Two Higgs Doublet Model with Flavor \newline Changing Currents}
\label{model}

A mild extension of the SM with one additional scalar SU(2) doublet
opens up the possibility of FCSC\null. For this reason, the 2HDM
scalar potential is usually constrained by an {\it ad hoc} discrete
symmetry \cite{glash}, whose only role is to protect the model from
tree-level FCSC. As a result one gets the so called Model I and Model
II, when up-type and down-type quarks are coupled to the same or to
two different doublets respectively \cite{hunter}. In particular, it
is to be stressed that from a purely phenomenological point of view,
low energy experiments involving $K^0$-$\bar K^0$, $B^0$-$\bar B^0$
mixing, $K_L\to\mu\bar\mu$ etc.\ place very stringent constraints only
on the existence of those tree level flavor changing transitions which
directly involve the first family. Indeed, in view of the
extraordinary mass scale of the top quark, it has been emphasized by
many that anomalously large flavor-changing (FC) couplings involving
the third family may exist \cite{sher}-\cite{mumutc},\cite{peccei}.
Thus, following Cheng and Sher \cite{sher}, perhaps a natural way to
limit the strength of the FCSC involving the first family is to assume
that they are proportional to the masses of the participating
quarks. In this way, the FC couplings are automatically put in a
hierarchical order and the third family may well play an enhanced
role.

For this type of 2HDM, the Yukawa Lagrangian for the quark fields can be
taken to have  the form \cite{savage,eetc}

\be
{\cal L}^{(III)}_{Y}= \eta^{U}_{ij} \bar Q_{i,L} \tilde\phi_1 U_{j,R} +
\eta^D_{ij} \bar Q_{i,L}\phi_1 D_{j,R} + 
\xi^{U}_{ij} \bar Q_{i,L}\tilde\phi_2 U_{j,R}
+\xi^D_{ij}\bar Q_{i,L} \phi_2 D_{j,R} \,+\, h.c. 
\label{lyukmod3}
\ee

\noindent where $\phi_i$, for $i=1,2$, are the two scalar doublets of
a 2HDM, while $\eta^{U,D}_{ij}$ and $\xi_{ij}^{U,D}$ are the non
diagonal coupling matrices. For convenience we can choose to express
$\phi_1$ and $\phi_2$ in a suitable basis such that only the
$\eta_{ij}^{U,D}$ couplings generate the fermion masses, i.e.\ such
that

\be
<\phi_1>=\left( 
\begin{array}{c}
0\\
{v/\sqrt{2}}
\end{array}
\right)\,\,\,\, , \,\,\,\,
<\phi_2>=0 
\ee 

\noindent The two doublets are in this case of the form

\be
\phi_1=\frac{1}{\sqrt{2}}\left\{\left(\ba{c} 0 \\ v+H^0\ea\right)+
\left(\ba{c} \sqrt{2}\,\chi^+\\ i\chi^0\ea\right)\right\}\,\,\,\,;\,\,\,\,
\phi_2=\frac{1}{\sqrt{2}}\left(\ba{c}\sqrt{2}\,H^+\\ H^1+i H^2\ea\right)
\ee

\noindent The scalar Lagrangian in the ($H^0$, $H^1$, $H^2$,
$H^{\pm}$) basis is such that\cite{knowles,hunter}:

\begin{enumerate}

\item the doublet $\phi_1$ corresponds to the scalar doublet of the SM
and $H^0$ to the SM Higgs field (same couplings and no interactions
with $H^1$ and $H^2$);

\item all the new scalar fields belong to the $\phi_2$ doublet;

\item both $H^1$ and $H^2$ do not have couplings to the gauge bosons
of the form $H^{1,2}ZZ$ or $H^{1,2}W^+W^-$.

\end{enumerate}

\noindent However, while $H^{\pm}$ is also the charged scalar mass
eigenstate, ($H^0$, $H^1$, $H^2$) are not the neutral mass
eigenstates. Let us denote by ($\bar H^0$, $h^0$) and $A^0$ the two
scalar plus one pseudoscalar neutral mass eigenstates. They are
obtained from ($H^0$, $H^1$, $H^2$) as follows

\bea 
\label{masseigen}
\bar H^0 & = & \left[(H^0-v)\cos\alpha + H^1\sin\alpha \right]
\nonumber \\ 
h^0 & = & \left[-(H^0-v)\sin\alpha + H^1\cos\alpha \right] \\ 
A^0 & = &  H^2
\nonumber 
\eea 

\noindent 
where $\alpha$ is a mixing angle, such that for $\alpha\!=\!0$, ($H^0$,
$H^1$, $H^2$) coincide with the mass eigenstates.  We find more
convenient to express $H^0$, $H^1$ and $H^2$ as functions of the mass
eigenstates, i.e.

\bea
\label{nomasseigen}
H^0 &=& \left(\bar H^0\cos\alpha-
h^0\sin\alpha\right)+v \nonumber \\
H^1 &=& \left( h^0\cos\alpha+\bar H^0\sin\alpha\right) \\
H^2 &=& A^0 \nonumber
\eea

\noindent In this way we may take advantage of the mentioned
properties (1), (2) and (3), as far as the calculation of the
contribution from new physics goes. In particular, only the $\phi_1$
doublet and the $\eta^U_{ij}$ and $\eta^D_{ij}$ couplings are involved
in the generation of the fermion masses, while $\phi_2$ is responsible
for the new couplings.

After the rotation that diagonalizes the mass matrix of the quark
fields, the FC part of the Yukawa Lagrangian looks like

\be
{\cal L}_{Y,FC}^{(III)} = \hat\xi^{U}_{ij} \bar Q_{i,L}\tilde\phi_2 
U_{j,R} +\hat\xi^D_{ij}\bar Q_{i,L} \phi_2 D_{j,R} \,+\, h.c. 
\label{lyukfc}
\ee

\noindent where $Q_{i,L}$, $U_{j,R}$ and $D_{j,R}$ denote now the
quark mass eigenstates and $\hat\xi_{ij}^{U,D}$ are the rotated
couplings, in general not diagonal. If we define $V_{L,R}^{U,D}$ to be
the rotation matrices acting on the up- and down-type quarks, with
left or right chirality respectively, then the neutral FC couplings
will be

\be
\hat\xi^{U,D}_{\rm neutral}=(V_L^{U,D})^{-1}\cdot \xi^{U,D}
\cdot V_R^{U,D}
\label{neutral}
\ee

\noindent On the other hand for the charged FC couplings we will have

\bea
\hat\xi^{U}_{\rm charged}\!&=&\!\hat\xi^{U}_{\rm neutral}\cdot 
V_{\sss{\rm CKM}}\nonumber\\
\hat\xi^{D}_{\rm charged}\!&=&\!V_{\sss{\rm CKM}}\cdot
\hat\xi^{D}_{\rm neutral} 
\label{charged}
\eea 

\noindent where $V_{\sss{\rm CKM}}$ denotes the
Cabibbo-Kobayashi-Maskawa matrix. To the extent that the definition of
the $\xi^{U,D}_{ij}$ couplings is arbitrary, we can take the rotated
couplings as the original ones. Thus, we will denote by
$\xi^{U,D}_{ij}$ the new rotated couplings in eq. (\ref{neutral}), such
that the charged couplings in (\ref{charged}) look like $\xi^{U}\cdot
V_{\sss{\rm CKM}}$ and $V_{\sss{\rm CKM}}\cdot\xi^{D}$.

We will assume that the $\xi^{U,D}_{ij}$ couplings are purely
phenomenological parameters and compare the region of the parameter
space that could accommodate $R^{\rm exp}_b$ with the constraints
from other physical processes. For convenience, we parametrize the
$\xi^{U,D}_{ij}$ couplings in such a way as to make the comparison
with the other 2HDM easier

\be
\xi^{U,D}_{ij}=\lambda_{ij}\,\frac{\sqrt{m_i m_j}}{v} 
\label{coupl_sher}
\ee 

\noindent This is very similar to what was proposed and used in
ref. \cite{sher,savage,eetc,mumutc}, but we want now to allow the
factors $\lambda_{ij}$ to vary over a broad range, constrained by
phenomenology only. In this way we may be able to see if the
experiment data lead to some new patterns in the coupling behavior
\cite{firstfamily}. 

\section{Implications for $R_b$ and $R_c$}
\label{rb}

Let us now focus on the calculation of $R_b$ and $R_c$. The main task
is to compute the corrections from new physics to the SM $Zq\bar q$
vertex, for $q=c,\,b$.  Suppose the reference SM vertex for a $Z\to
q\bar q$ process is

\be
V_{q\bar q Z}^{\sss{\rm SM}}\equiv -i\frac{g_{\sss W}}{c_{\sss W}}\bar 
q\gamma_{\mu}\left[\Delta_{q,L}^{\sss{\rm SM}}\frac{(1-\gamma_5)}{2}+
\Delta_{q,R}^{\sss{\rm SM}}\frac{(1+\gamma_5)}{2}\right]q Z^{\mu}
\label{vert_qSM}
\ee

\noindent where $c_{\sss W}$ is the cosine of the Weinberg angle and
$g_{\sss W}$ is the weak gauge coupling. The presence of new
interactions will then modify it into

\be
V_{q\bar q Z}\equiv -i\frac{g_{\sss W}}{c_{\sss W}}\bar q\gamma_{\mu}
\left[\Delta_{q,L}\frac{(1-\gamma_5)}{2}+
\Delta_{q,R}\frac{(1+\gamma_5)}{2}\right]q Z^{\mu}
\label{vert_qtot}
\ee

\noindent where

\be
\Delta_{q,L(R)}\equiv\Delta_{q,L(R)}^{\sss{\rm SM}}+
\Delta_{q,L(R)}^{\sss{\rm NEW}}
\label{deltadef}
\ee

\noindent is the sum of the original SM contribution plus the new one
from the $\xi$-type scalar couplings. In principle, both SM and Model
III radiative corrections to the $Zq\bar q$ vertex give origin to one
additional form factor, proportional to $\sigma^{\mu\nu}q_{\nu}$ (the
$\sigma^{\mu\nu}q_\nu\gamma_5$ form factor is absent because it would
violate CP).  This magnetic moment-type form factor arises at one-loop
and should be considered as well.  We have calculated it and verified
that, as is the case in the SM, it is very small, at least three
orders of magnitude smaller than the leading contributions to
$\Delta_{q,L(R)}^{\sss{\rm NEW}}$. Therefore, we neglect its effect
in the following discussion. 

In view of the previous discussion and neglecting all finite quark
mass effects ($m_q\sim 0$) \cite{massless}, the generic expression for
$\Gamma(Z\to q\bar q)$, for $q=b,\,c$, can then be written as

\be \Gamma(Z\to q\bar q)=\frac{N_c}{6}\frac{\hat\alpha}
{{\hat s}_{\sss W}^2 {\hat c}_{\sss W}^2}M_Z
\left((\Delta_{q,L})^2+(\Delta_{q,R})^2\right)
\label{gammaqq}
\ee

\noindent where all kinds of EW+QCD corrections have been reabsorbed
in the redefinition of the QED fine-structure constant $\alpha$, of
$c_{\sss W}$ ($s_{\sss W}$) and of the couplings
$\Delta_{q,L(R)}$. Moreover, the $\Delta_{q,L(R)}$ couplings contain
corrections induced by the new FC scalar couplings. 

In order to compute the corrections to $R_q$ from new physics, such as
due to the scalar fields of Model III, we observe that, since $R_q$ is
the ratio between two hadronic widths, most EW oblique and QCD
corrections cancel, in the massless limit, between the numerator and
the denominator. The remaining ones are absorbed in the definition of
the renormalized couplings $\hat\alpha$ and ${\hat s}_{\sss W}$
(${\hat c}_{\sss W}$), up to terms of higher order in the electroweak
corrections \cite{pich,kuhn,grant}. As a consequence, the
$\Delta_{q,L(R)}$ couplings will be as in eq. (\ref{vert_qtot}), with
$\Delta_{q,L(R)}^{\sss{\rm SM}}$ given by the tree level SM couplings
expressed in terms of the renormalized couplings $\hat\alpha$ and
${\hat s}_{\sss W}$ (${\hat c}_{\sss W}$).  This feature makes the
study of $R_b$ and $R_c$ particularly interesting, because the new FC
contributions may be easily disentangled in the $Zq\bar q$-vertex
corrections. In fact, the presence of new scalar-fermion couplings
will affect the $W$ and $Z$ renormalized propagators too, giving
stringent constraints especially from the corrections to the $\rho$
parameter. However, this is not relevant for the specific calculation
of $R_b$ and will be discussed in later segments of this paper.

In light of the preceding remarks, we can express $R_b$ and $R_c$ in
terms of $R_b^{\sss{\rm SM}}$ and $R_c^{\sss{\rm SM}}$ as follows:

\be
R_q  =  R_q^{\sss{\rm SM}} \frac{1+\delta_q}{[1+R^{\sss{\rm SM}}_b
\delta_b+R^{\sss{\rm SM}}_c\delta_c]} \label{rqrsmb}
\ee

\noindent where

\be
\delta_q = 2\,\frac{\Delta^{\sss{\rm SM}}_{qL} \Delta^{\sss{\rm
NEW}}_{qL} + \Delta^{\sss{\rm SM}}_{qR} \Delta^{\sss{\rm
NEW}}_{qR}}{(\Delta^{\sss{\rm SM}}_{qL})^2 + ( \Delta^{\sss{\rm
SM}}_{qR})^2} \label{alphaq}
\ee

\noindent for $q=b,c$. In eq. (\ref{rqrsmb}), terms of
$O((\Delta^{\sss{\rm NEW}}_{qL(R)})^2)$ have been neglected and the
numerical analysis confirms the validity of this approximation.
\begin{figure}
\centering
\epsfxsize=5.in
\leavevmode\epsffile{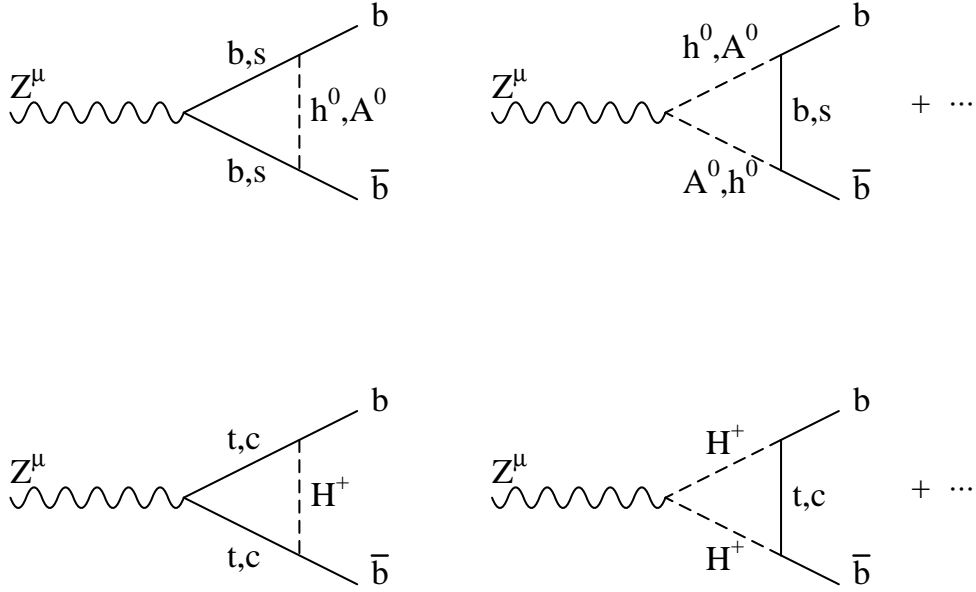}
\caption[]{Typical corrections to the $Z b\bar b$ vertex due to both 
charged and neutral scalar/pseudoscalar, in Model III.}
\label{fd_Zbb}
\end{figure}

In particular, we will have to compute $\Delta^{\sss{\rm
NEW}}_{b,L(R)}$ and $\Delta^{\sss{\rm NEW}}_{c,L(R)}$ in our model. In
Fig. \ref{fd_Zbb} we show a sample of the Feynman diagrams which
correspond to the corrections to the $Zb\bar b$ vertex, due to both
charged and neutral scalars/pseu\-do\-sca\-lars. The $Z c\bar c$ case
is strictly analogous, up to modifications of the external and
internal quark states. In our calculation, we will assume that the FC
couplings involving the first generation are negligible and we will
consider all the other possible contributions from the new $\xi$-type
vertices, containing both flavor-changing and flavor-diagonal terms
(see eqs. (\ref{lyukfc})--(\ref{coupl_sher})).

We examined all the possible scenarios, varying the scalar
masses ($M_H$, $M_h$, $M_A$ and $M_c$), the mixing angle ($\alpha$)
and  the $\xi$-couplings. The striking result emerging from
this analysis is that, in spite of the arbitrariness of the new FC
couplings, there exists only a very tight window in which the
corrections from this new physics enhance $R_b$, to make it compatible
with the experimental indications.  We find maximum enhancement for 
\begin{itemize}
\item very large $h^0 b\bar b$ and $A^0 b\bar b$ couplings, obtained for
\be
\xi^D_{bb}\ge 60\frac{m_b}{v}\; ;
\label{csibb}
\ee

\item the phase $\alpha=0$;
\item light and approximately equal neutral scalar and pseudoscalar 
masses: $M_h\sim M_A\sim 50$ GeV (i.e.\ at the edge of the allowed
experimental lower bound for $M_h$ and $M_A$ \cite{pdg});

\item much heavier charged scalar masses, i.e.\ $M_c\sim 400$ GeV or
more. Lighter charged masses require even more demanding bounds on the
previous parameters.
\end{itemize}
For these values of the parameters we can get:
\be
0.2185\leq R_b\leq 0.2230 \label{rb_mod3}
\ee

\noindent i.e.\ quite consistent with the experimental measurements,
$R^{\rm exp}_b = .2219\pm .0017$ \cite{zbbzcc}.

We note that the enhanced coupling (\ref{csibb}) to the $b$ quark means
$\xi^D_{bb}\sim \xi^U_{tt}$ (with $\lambda_{tt}\sim 1$). Perhaps this
signifies the special role of the third family with respect to Higgs
interactions. For our purpose, of course, these couplings are purely
phenomenological.

The previous set of parameters strictly mimic what was already found
in the context of Model II, i.e.\ without tree-level FCNC\null. Indeed
our model can be compared to that one when the phase $\alpha=0$, and
the FC couplings are set to zero. In this regime, we confirm the
results of Ref. \cite{kuhn,grant}. The pattern of cancellation between
neutral and charged contributions is still valid in Model III as
well. The charged contribution to $\Delta^{\sss{\rm NEW}}_{b,L(R)}$ is
negative and tends to reduce $R_b$, while the neutral one, for light
scalar masses ($M_{h,A}\le 100$ GeV), is positive and tends to enhance
$R_b$. With an assumption like the one in eq. (\ref{coupl_sher}), the
neutral scalar and pseudoscalar vertex corrections are suppressed due
to their small couplings to the $b$-quark, unless
$\lambda_{bb}\gg1$. Thus in order to enforce the cancellation, we have
to enhance these couplings as in eq. (\ref{csibb}) as well as to demand
the charged scalar to be much heavier than the neutral scalar and
pseudoscalar.

The crucial difference between the two models is that Model III,
unlike Model II, does not provide any relation between $\xi^U$- and
$\xi^D$-type couplings. In fact, for $\xi^D_{bb}\sim 60\, m_b/v$ as in
eq. (\ref{csibb}), we have that $\xi^D_{bb}\sim \xi^U_{tt}$, while in
Model II $\xi^D_{bb}$ would be inversely proportional to $\xi^U_{tt}$
and we would have at the same time a very enhanced $\xi^D$-type
coupling and a very suppressed $\xi^U$-type one. This is at the origin
of the slightly more demanding bounds we have to impose on the
parameters of Model III with respect to Model II if we want
$R_b>R_b^{\sss{\rm SM}}$. This difference will become even more
important in the discussion of the other constraints, as we will see
in a while.

Moreover, in Model III there are also FC couplings, such as
$\xi^D_{sb}$ and $\xi^U_{ct}$. We note that, as far as $R_b$ is
concerned, $\xi^U_{ct}$ plays a role only in the charged contribution
to $\Delta^{\sss{\rm NEW}}_{b,L(R)}$ and, since this contribution is
negative, we do not want to enhance it. On the other hand,
$\xi^D_{sb}$ affects both the neutral and the charged vertex diagram,
thus, in principle, it could play some role. However, even with any
reasonable enhancement, $\xi^D_{sb}$ does not seem to change the
result significantly.

The scenario we find turns out to be greatly modified when we
incorporate two additional constraints: the correction to the $\rho$
parameter, and the implication for $Br(B\to X_s\gamma)$. In fact, in
the framework of Model III with enhanced $\xi^D_{bb}$ coupling, the
first one turns out to be very sensitive to a heavy $M_c$, while the
second imposes a severe restriction on the magnitude of the
$\xi^D_{bb}$ coupling. Let us illustrate them in turn.

\section{$\rho$-Parameter  Constraints on Model III}
\label{rho}

The relation between $M_W$ and $M_Z$ is modified by the presence of
new physics and the deviation from the SM prediction is usually
described by introducing the parameter $\rho_0$ \cite{pdg,lang},
defined as

\be
\rho_0=\frac{M_W^2}{\rho M_Z^2\cos^2\theta_W}
\label{rhozero_def}
\ee

\noindent where the $\rho$ parameter reabsorbs all the SM corrections
to the gauge boson self-energies. We recall that the most important SM
corrections at the one-loop level are induced by the top quark
\cite{grant,lang}

\be
\rho_{\rm top}\simeq\frac{3 G_F m_t^2}{8\sqrt{2}\pi^2}
\label{rhotop}
\ee

\noindent Within the SM with only one scalar SU(2) doublet
$\rho_0^{\rm tree}=1$. In the presence of new physics we have

\be
\rho_0=1+\Delta\rho_0^{\sss{\rm NEW}} 
\label{rhozero}
\ee

\noindent where $\Delta\rho_0^{\sss{\rm NEW}}$ can be written in terms
of the new contributions to the $W$ and $Z$ self-energies as

\be
\Delta\rho_0^{\sss{\rm NEW}}=\frac{A^{\sss{\rm NEW}}_{WW}(0)}{M_W^2}-
\frac{A^{\sss{\rm NEW}}_{ZZ}(0)}{M_Z^2}
\ee

\noindent Using the general analytical expressions in ref. \cite{bert},
and adapting the discussion to Model III (making use of the Feynman
rules given in Appendix \ref{fr_mod3}), we find that

\be
\Delta\rho_0^{\sss{\rm NEW}}\simeq\frac{G_F}{8\sqrt{2}\pi^2}
\left(\sin^2\!\alpha\,
G(M_c,M_A,M_H)+\cos^2\!\alpha\, G(M_c,M_A,M_h)\right)
\label{deltarhozero}
\ee

\noindent where all the terms of order $(M_{W,Z}^2/M_c^2)$ have been
neglected and we define

\bea
& &G(M_c,M_A,M_{H,h})= M_c^2-
\frac{M_c^2 M_A^2}{M_c^2-M_A^2}\log{\frac{M_c^2}{M_A^2}}
\nonumber\\
&-&\!\frac{M_c^2 M_{H,h}^2}{M_c^2-M_{H,h}^2}\log{\frac{M_c^2}{M_{H,h}^2}}
+\frac{M_A^2 M_{H,h}^2}{M_A^2-M_{H,h}^2}
\log{\frac{M_A^2}{M_{H,h}^2}}
\label{gfunction}
\eea

\noindent The determination of $m_t$ from FNAL \cite{cdf} allows us to
distinguish between $\rho_0$ and $\rho\simeq 1+\rho_{\rm top}$. {}From
the recent global fits of the electroweak data, which include the
input for $m_t$ from ref. \cite{cdf} and the new results on $R_b$,
$\rho_0$ turns out to be very close to unity. For $R_b$=$R^{\rm
exp}_b$ as in eq. (\ref{rbexp}) and $m_t=(174\pm 16)$ GeV,
ref. \cite{lang} quotes

\be
\rho_0=1.0004\pm 0.0018\pm 0.0018
\label{rhozero_gf}
\ee

This result clearly imposes stringent limits on the parameters of any
extended model. In particular, if we refer to Section \ref{rb} and
evaluate $\Delta\rho_0^{\sss{\rm NEW}}$ for the set of parameters
which was found to give an enhanced value of $R_b$, we find that

\be
\Delta\rho^{\sss{\rm NEW}}\simeq \frac{G_F}{8\sqrt{2}\pi^2}M_c^2 
\label{deltarho}
\ee

\noindent where the neglected terms are suppressed as
$(M_{h,A}^2/M_c^2)$ or $(M_{W,Z}^2/M_c^2)$. We observe that, for
$\alpha\!=\!0$, the contributions of the $\phi_1$ and $\phi_2$
doublets are completely decoupled and the new physics contributions
come from the $\phi_2$ doublet only. The $\phi_1$ doublet can indeed
be identified with the usual SM Higgs doublet and its contribution to
$\rho_0$ is already included in the SM value of $\rho$. Using
eq. (\ref{deltarho}), eqs. (\ref{rhozero}) and (\ref{rhozero_gf}) lead
to the following upper bound on the charged scalar mass

\be
M_c\leq 200\, \mbox{GeV} 
\label{mclsim}
\ee

\noindent The upper bound (\ref{mclsim}) for $M_c$ means that, to
retain $R_b$ in the range of eq. (\ref{rb_mod3}) would require even
larger coupling $\xi^D_{bb}$ than in eq. (\ref{csibb}), since the
latter was obtained with $M_c\ge400$ GeV and since, also, we cannot
reduce the neutral scalar masses below their experimental bounds.

\section{Implications of $b\to s\gamma$}
\label{bsgamma}

Even more dramatically, the requirement of an enhanced $\xi^D_{bb}$
coupling clashes with the experimental constraint for $Br(B\to
X_s\gamma)$ \cite{alam}

\be
Br(B\to X_s\gamma)=(2.32\pm 0.51\pm 0.29\pm 0.32)\times 10^{-4}
\ee

\noindent where the first error is statistical and the latter two are
systematic errors.

This is a remarkable difference with respect to other 2HDM, in which
there is still a small compatibility between an enhancement over
$R_b^{\sss{\rm SM}}$ and the result for $Br(B\to X_s\gamma)$ obtained
by the experiment \cite{grant}. We will not consider Model I, because
it cannot produce an acceptable answer for $R_b$, since the
fermion-scalar couplings in this model are either all simultaneously
enhanced or simultaneously suppressed. Thus a disparity between
neutral and charged scalar vertex corrections can never be realized in
Model I. Instead, let us focus on Model II and Model III\null. It is
interesting to compare what ``the enhancement of the $\xi^D_{bb}$
coupling'' means in these two models. We then immediately realize that
in Model III this implies a new large contribution from the neutral
scalar and pseudoscalar penguin diagrams and an enormous enhancement
of the charged scalar penguin diagram, due to the link between neutral
and charged coupling via eq. (\ref{charged}).

To calculate the contribution of $h_0$, $A_0$ and $H^{\pm}$ to the
$Br(B\to X_s\gamma)$, we work in the effective Hamiltonian formalism,
thereby including also QCD corrections at the leading order
\cite{alpha0}. Due to the presence of new effective interactions, we
need to modify both the basis of local operators in the effective
Hamiltonian and the initial conditions for the evolution of the Wilson
coefficients.  This is a well known procedure for calculating the
effect of heavy new degrees of freedom which do not appear in the
evolution of the coefficients at low energy, but only in their initial
conditions at an initial scale roughly set at $\mu\sim M_W$. We refer
to the literature for all the necessary technical details
\cite{bsg_all,ciuetal1,buras}.

In particular, when we include the new heavy degrees of freedom
($h_0$, $A_0$ and $H^{\pm}$), there are two main changes that we need
to consider.  First, there are now two QED magnetic-type operators
with opposite chirality, which we denote by $Q_7^{(R,L)}$ and write as
\cite{normal}

\be
Q_7^{(R,L)}=\frac{e}{8\pi^2}m_b\bar s\sigma^{\mu\nu}(1\pm\gamma_5)b 
F_{\mu\nu}
\ee

\noindent We recall that in the SM as well as in Model II 
the absence of $Q_7^{(L)}$ is a consequence of assuming $m_s/m_b\sim
0$.  In Model III, we do not want to make any {\it a priori}
assumption on the $\xi$-couplings, because of their arbitrariness, and
therefore both $Q_7^{(R)}$ and $Q_7^{(L)}$ can contribute to the $b\to
s\gamma$ decay. The rate $\Gamma(b\to s\gamma)$ will be proportional
to the sum of the modulus square of their coefficients at a scale
$\mu\sim m_b$, i.e.

\be
\Gamma(b\to s\gamma)
\sim\left(|C_7^{(R)}(m_b)|^2+|C_7^{(L)}(m_b)|^2\right)
\ee

\noindent We observe that, due to their opposite chirality, the two
operators $Q_7^{(R,L)}$ do not mix under QCD corrections and, in a
first approximation, their evolution with the scale $\mu$ can be taken
to be the same as in the SM (for $Q_7^{(R)}$) and equal for both of
them. In so doing, we neglect those operators whose effect is
sub-leading either because of their chiral structure or because of the
heavy mass of the scalar boson which generates them.

The second change concerns the initial conditions for the Wilson
coefficients at a scale $\mu\sim M_W$. $C_7^{(R,L)}(m_b)$ depend in
general on many initial conditions. However, for the same reasons
explained before, the most relevant new contributions, due both to
neutral and charged scalar fields, mainly affect
$C_7^{(R,L)}(M_W)$. In the following we will discuss the results of
our numerical evaluation of both neutral and charged contributions and
their impact on the decay rate for $b\to s\gamma$. In particular, we
will focus on the rate normalized to the QCD corrected semileptonic
rate, i.e.\ on the ratio:

\bea
R&=&\frac{\Gamma(B\to X_s\gamma)}
{\Gamma(B\to X_c e\bar\nu_e)}\sim
\frac{\Gamma(b\to s\gamma)}
{\Gamma(b\to ce\bar\nu_e)} \nonumber \\
&=& \frac{6\alpha}{\pi\,f(m_c/m_b)}\, F\,
\left(|C_7^{(R)}(m_b)|^2+|C_7^{(L)}(m_b)|^2\right)
\label{ratio}   
\eea

\noindent where $f(m_c/m_b)$ is the phase-space factor for the
semileptonic decay and $F$ takes into account some $O(\alpha_s)$
corrections to both $B\rightarrow X_c e\bar\nu_e$ and $B\rightarrow
X_s\gamma$ decays (see ref. \cite{ciuetal2} for further comments). We
also neglect possible deviations from the spectator model prediction
of $\Gamma(B\rightarrow X_s\gamma)$ and $\Gamma(B\to
X_ce\bar\nu_e)$. {}From eq. (\ref{ratio}) a convenient theoretical
prediction for $Br(B\to X_s\gamma)$ can be extracted, to be compared
with the experimental result.

As far as the new FC contributions from neutral scalar and pseudoscalar
go, they are peculiar to Model III, because they contain FC couplings.
Were it not for the enhancement of $\xi^D_{bb}$, they would be
completely negligible. When $\xi^D_{bb}\ge 60 m_b/v$ however, the
$h_0$ and $A_0$ penguin diagrams give a sizable contribution,
amounting to about 30\% correction to the SM amplitude. This is still
within the range allowed by the experiments, and constitute a first
non-negligible point of difference with respect to Model II\null.

However, the most striking effect emerges when we consider
the charged scalar penguin. Let us focus separately on
$C_7^{(R)}(M_W)$ and $C_7^{(L)}(M_W)$ and try to make a direct
comparison with Model II\null. We recall that the charged couplings for
Model II are given by

\be
{\cal L}_{Y}^{(II)}=\sqrt{\frac{4G_F}{\sqrt{2}}}H^+\left[
\tan{\beta}\,\bar U_L V_{\sss{\rm CKM}} M_D D_R + 
\frac{1}{\tan{\beta}}\,\bar U_R M_U V_{\sss{\rm CKM}} D_L\right] + h.c.
\label{lyuk2}
\ee

\noindent where $M_U$ and $M_D$ are the diagonal mass matrices for the
U-type and D-type quarks respectively, and $\tan{\beta}=v_2/v_1$ is
the ratio between the vacuum expectation values of the two scalar
doublets. The analogous couplings for Model III are expressed by
eqs. (\ref{lyukfc}) and (\ref{charged}).

Both in Model II and in Model III, the new contributions to
$C_7^{(R)}(M_W)$ happens to be multiplied by two products of Yukawa
couplings, which we will denote by $(\xi^{U*}_{st}\xi^U_{tb})_{\rm
ch}$ and $(\xi^{U*}_{st}\xi^D_{tb})_{\rm ch}$.  Using
eq. (\ref{lyuk2}), we derive that, in Model II these products of
Yukawa couplings are given by

\bea
(\xi^{U*}_{st}\xi^U_{tb})_{\rm ch}^{(II)}\!&=&\!
\frac{4G_F}{\sqrt{2}}V^*_{ts}V_{tb} m_t^2\frac{1}{\tan{\beta}^2} 
\nonumber\\
(\xi^{U*}_{st}\xi^D_{tb})_{\rm ch}^{(II)}\!&=&\!
\frac{4G_F}{\sqrt{2}}V^*_{ts}V_{tb} m_t m_b
\label{coupl_mod2}
\eea

\noindent On the other hand, in Model III, using eqs. (\ref{lyukfc})
and (\ref{charged}) they can be written as

\bea
(\xi^{U*}_{st}\xi^U_{tb})_{\rm ch}^{(III)}\!&=&\! V^*_{ts}V_{tb}
\left(\xi^U_{tt}+\xi^U_{ct}\frac{V^*_{cs}}{V^*_{ts}}\right)
\left(\xi^U_{tt}+\xi^U_{tc}\frac{V_{cb}}{V_{tb}}\right) \nonumber\\
(\xi^{U*}_{st}\xi^D_{tb})_{\rm ch}^{(III)}\!&=&\! V^*_{ts}V_{tb}
\left(\xi^U_{tt}+\xi^U_{ct}\frac{V^*_{cs}}{V^*_{ts}}\right)
\left(\xi^D_{bb}+\frac{V_{ts}}{V_{tb}}\xi^D_{sb}\right)
\label{coupl_mod3}
\eea
 
\noindent In order to compare the two models, let us use the
parameterization introduced in eq. (\ref{coupl_sher}) and let us set
all the FC couplings in Model III to zero, namely $\xi^U_{ct}\!=\!0$
and $\xi^D_{sb}\!=\!0$. Then, the couplings in eq. (\ref{coupl_mod3})
reduce to the following form:

\bea
(\xi^{U*}_{st}\xi^U_{tb})_{\rm ch}^{(III)}\!&=&\!V^*_{ts}V_{tb}
(\xi^U_{tt})^2=
\frac{4G_F}{\sqrt{2}}V^*_{ts}V_{tb}(\lambda_{tt})^2 m_t^2 \nonumber\\
(\xi^{U*}_{st}\xi^D_{tb})_{\rm ch}^{(III)}\!&=&\!V^*_{ts}V_{tb}
\xi^U_{tt}\xi^D_{bb}=
\frac{4G_F}{\sqrt{2}}V^*_{ts}V_{tb} (\lambda_{tt}\lambda_{bb}) m_t m_b
\label{coupl_mod3two}
\eea

\noindent {}From eqs. (\ref{coupl_mod2}) and (\ref{coupl_mod3two}), 
the different behavior of Model II and Model III with respect to an
enhancement of the $\xi^D_{bb}$-like coupling should be clear. The
following correspondence holds:

\be \ba{ccccc}
\phantom{(\xi^{U*}_{st}\xi^U_{tb})_{\rm ch}}&\phantom{ : }&
\mbox{Model II} & \phantom{\to} & \mbox{Model III} \\ \\
(\xi^{U*}_{st}\xi^U_{tb})_{\rm ch}& : & \frac{1}{\tan{\beta}^2} &
\to & \lambda_{tt}^2 \\ (\xi^{U*}_{st}\xi^D_{tb})_{\rm ch} &
: & 1 & \to & \lambda_{tt}\lambda_{bb}
\label{coupl_comp}
\ea
\ee

\noindent In Model II, the enhancement of $\xi^D_{bb}$ corresponds to
the choice of large value for $\tan{\beta}$, i.e.\ to a suppression of
the $(\xi^{U*}_{st}\xi^U_{tb})_{\rm ch}$ coupling with respect to the
$(\xi^{U*}_{st}\xi^D_{tb})_{\rm ch}$ one, which stays the same, i.e.\
pretty small. In Model III, on the other hand, we just require
$\lambda_{bb}\ge 60$ to enhance $R_b$, but we do not have any reason
to reduce $\lambda_{tt}$, since each coupling is independent and
arbitrary.  As a net result the charged scalar penguin diagram is
greatly enhanced in Model III, even with $\xi^U_{ct}=0$ and
$\xi^D_{sb}=0$. If we restate these FC couplings to their non-zero
value, the situation is even worse.

Let us now consider $C_7^{(L)}(M_W)$. This coefficient is special to
Model III since it is normally neglected in Model II in the limit
$m_s/m_b\sim 0$. It turns out to be proportional to the other two
possible combinations of Yukawa couplings, i.e.

\bea
(\xi^{D*}_{st}\xi^U_{tb})_{\rm ch}^{(III)}\!&=&\! V^*_{ts}V_{tb}
\left(\frac{V_{tb}}{V^*_{ts}}\xi^D_{bs}+\xi^D_{ss}\right)
\left(\xi^U_{tt}+\xi^U_{tc}\frac{V_{cb}}{V_{tb}}\right) \nonumber\\
(\xi^{D*}_{st}\xi^D_{tb})_{\rm ch}^{(III)}\!&=&\! V^*_{ts}V_{tb}
\left(\frac{V_{tb}}{V^*_{ts}}\xi^D_{bs}+\xi^D_{ss}\right)
\left(\xi^D_{bb}+\frac{V_{ts}}{V_{tb}}\xi^D_{sb}\right)
\label{coupl_mod3three}
\eea

\noindent and constitutes a relevant extra contribution to
$Br(B\to X_s\gamma)$, to the extent that the FC couplings,
namely $\xi^D_{bs}$ and $\xi^U_{ct}$, are not negligible. 

{}From a numerical analysis, we obtain that for $M_c=200$ GeV, Model
III contribution is about a factor of 40 larger than the SM
amplitude. When $M_c$ increases to about 3-4 TeV the two contributions
become comparable. Thus $Br(B\to X_s\gamma)$ restricts

\be
M_c\gsim 5\mbox{ TeV} \label{mcgsim}
\ee

\noindent in this version of Model III with the enhanced coupling of
eq. (\ref{csibb}) that is needed to account for $R_b$.

Since the Model II prediction for $Br(B\to X_s\gamma)$ is already
barely compatible with experiment, unless $M_c$ is quite big, the
previous comparison clearly shows that any enhancement of the
$\xi^D_{bb}$ coupling, i.e.\ of $R_b$, cannot be accommodated by Model
III\null.

\section{Remarks on the Experimental Aspects of $R_b$ and $R_c$;
$R_{b+c}$ and $R_\ell$.}
\label{exp}

The preceding discussion leads us to conclude that Model III cannot
simultaneously satisfy the constraints from the $\rho$-parameter,
$Br(B\to X_s\gamma)$ and $R^{\rm exp}_b$. Therefore, the model may
well be wrong and/or incomplete.  We view the model as an illustration
of the kind of theoretical scenarios that can result from a rather
minimal extension of the SM, namely due to the introduction of an
extra Higgs doublet.  The main virtue of the model is that it gives a
reasonably well defined theoretical framework in which experimental
constraints on flavor-changing-scalar couplings can be systematically
categorized.

While the model may well be wrong, it is perhaps also of some use to
question the experimental results i.e.\ $R^{\rm exp}_b$ (and $R^{\rm
exp}_c$). As alluded to in the Introduction, the experimental analysis
for $R_b$ and $R_c$ are correlated \cite{eps}. The deviation from the
SM given in eq. (\ref{rbexp}) appears quite significant
($\sim3\sigma$), but this is only after the results from all the four
LEP detectors, and several different data sets are combined, including
their systematic errors.  One interesting aspect of the $R_b$ results
is that all the experiments find that $R^{\rm exp}_b> R^{\sss{\rm
SM}}_b$, although the significance of individual data sets is
typically${}\sim{}$(1--2)$\sigma$. The final errors given in
eq. (\ref{rbexp}) include statistical and systematic errors. To the
extent that the experiments are truly independent, one is tempted to
interpret that they are confirming each other at least on this overall
trend. On the other hand, it is also conceivable that this is a
reflection of the fact that some of the systematics (shared by the
experiments) are causing the problem.

Ironically $R^{\rm exp}_b$ and $R^{\rm exp}_c$ deviate oppositely
from the SM values. In fact, using ref. \cite{eps} we get

\bea
R^{\rm exp}_b + R^{\rm exp}_c & = & (.2219 \pm .0017) + (.1543\pm
.0074) \nonumber \\
& = & .376\pm.018 \label{rbcexp}
\eea

\noindent which is quite consistent with the SM

\be 
R^{\sss{\rm SM}}_b + R^{\sss{\rm SM}}_c = .388 \label{rbcsm}
\ee

\noindent It is then natural to be concerned that the experimental
effect could, in part, arise from misidentification of flavors.

Indeed $R_{b+c}$ defined as

\be
R_{b+c} = \frac{\Gamma(Z\to b\bar b \mbox{ or } c\bar c)}{\Gamma
(Z\to \mbox{hadrons})} \label{rbpc}
\ee

\noindent is a very useful observable. It shares the theoretical
cleanliness of $R_b$ and $R_c$: it is insensitive to QCD corrections.
It has significant experimental advantages, though, as separation
between $b$ and $c$ (which is often difficult) need not be made. As a
specific example, when charm or bottom decay semi-leptonically, the
hardness of the lepton is often used to distinguish bottom from charm.
With the use of $R_{b+c}$, one only needs to separate these heavy
flavors from the really light ones ($u,d,s$).

Of course $R^{\rm exp}_{b+c}$ cannot be obtained by adding the
existing numbers for $R^{\rm exp}_b$ and $R^{\rm exp}_c$ and we will
have to await a separate experimental analysis for that. Meantime, we
note that $R_\ell$ given by

\be
R_\ell = \frac{\Gamma(Z\to\mbox{hadrons})}{\Gamma(Z\to \ell^+\ell^-)}
\label{relleq}
\ee

\noindent for which experimental numbers are available \cite{eps} does
contain information on $\Gamma(Z\to b\bar b \mbox{ or } c\bar c)$. 
Indeed \cite{eps}

\be
R^{\rm exp}_\ell = 20.788\pm.032 \label{rexpell}
\ee

\noindent is rather precisely known with an accuracy of${}\sim.15\%$
which is significantly better than $R^{\rm exp}_b$ (0.7\%) or $R^{\rm
exp}_c$ (4.5\%). $R_\ell$, though, does depend on QCD corrections.
The calculation of $R_\ell$ is outlined in Appendix \ref{Rl}\null. 

It is important to observe that, to calculate the SM prediction
($R^{\sss{\rm SM}}_\ell$) we need to use $\alpha_s(M_Z)$ deduced from
other physical methods (i.e.\ not $\Gamma(Z\to{}$hadrons)). In this
way, $R^{\rm exp}_\ell$ can provide another constraint on any
global fit of the SM.  Two independent determinations of
$\alpha_s(M_Z)$, for example, come from the lattice \cite{latt,pdg}
and from the event shapes in $e^+e^-$ annihilation \cite{pdg}

\bea
\alpha^{\rm latt}_s (M_Z) & = & .110 \pm .006 \nonumber \\
\alpha^{e^+e^-}_s(M_Z) & = & .121\pm.006 \label{alphalatt}
\eea

\noindent We will use the average of the two: $\bar\alpha_s(M_Z)
\simeq .116\pm.006$. Using Table \ref{Delta_QCD} in Appendix \ref{Rl},
we then get the SM prediction

\be
R^{\sss{\rm SM}}_\ell = 20.748\pm.043 \label{rsmell}
\ee

\noindent The error in eq. (\ref{rsmell}) corresponds to the .006
error (to $1\sigma$) estimates on the central value of
$\bar\alpha_s(M_Z)$. Comparing eqs. (\ref{rexpell}) and (\ref{rsmell}),
we see that $R^{\sss{\rm SM}}_\ell$ is consistent with the
experimental number, i.e. within about $1\sigma$ of the error on the
experiment alone.

In passing we note that if the true $\alpha_s(M_Z)$ was taken to be
0.110 then 

\be
R_\ell [\alpha_s(M_Z) = 0.110] = 20.706 \label{rellalpha}
\ee

\noindent which would start to deviate from the experimental result in
eq. (\ref{rexpell}) at the $2.6\sigma$ level. But, with the current
experimental accuracy, this deviation only occurs if one attributes
essentially no error to the .110 central value of
$\alpha_s(M_Z)$\cite{shifman}. We do not consider it reliable, at
present, to reduce the theoretical errors so sharply. It is clearly
important, though, that the efforts towards improved evaluations of
$\alpha_s(M_Z)$ be continued, as then the experimental precision on
$R_\ell$ could be used more effectively to signal new physics.

\section{Disregarding $R^{\rm exp}_b$}
\label{norb}

Given the previous analysis, we want now to reexamine Model III
without imposing the constraint coming from $R_b^{\rm exp}$. Instead,
we will give predictions for $R_b$, $R_c$ and $R_{b+c}$ from the
model, subjecting it only to the $\rho$-parameter and $Br(B\rightarrow
X_s\gamma)$.
\begin{figure}
\centering
\epsfxsize=4.in
\leavevmode\epsffile{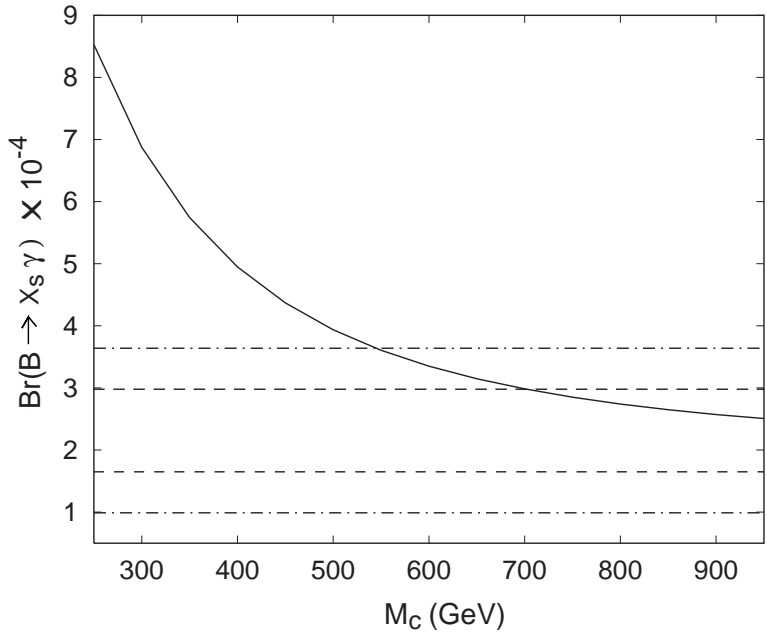}
\caption[]{$Br(B\to X_s\gamma)$ in Model III. The experimental result 
at $1\sigma$ (dashed) and $2\sigma$ (dot-dashed) is also given.}
\label{bsg}
\end{figure}

If we disregard $R_b^{\rm exp}$, then there is no need to enhance
$\xi^D_{bb}$ and we can take $\lambda_{bb}=1$ in
eq. (\ref{coupl_sher}). In this case, Model III predicts a $Br(B\to
X_s\gamma)$ compatible with experiments at the $2\sigma$-level, for
$M_c\ge 600$ GeV, as we can see in Fig. \ref{bsg}. As soon as
$\xi^D_{bb}$ is not enhanced anymore, the contribution of the neutral
scalars and pseudoscalar is completely negligible. Therefore, both the
value of the mixing angle $\alpha$ and of the neutral scalar and
pseudoscalar masses ($M_{H}$, $M_h$ and $M_A$) are irrelevant. In
particular, Fig. \ref{bsg} is obtained for $\alpha=\pi/4$ and values
for ($M_H$, $M_h$, $M_A$) resulting from the fit to $\Delta\rho_0$, as
we will discuss in a while. Due to the qualitative character of our
analysis, at this point it sufficies to seek consistency with the
experiment at the $2\sigma$-level. Indeed, we took as reference the SM
calculation \cite{ciuetal2}, which is already affected by a large
uncertainty, and computed only the leading corrections due to the new
scalar bosons of Model III, i.e. without considering the complete LO
effective hamiltonian analysis. {}From Fig. \ref{bsg} we also note
that, for $M_c\ge 600$ GeV, Model III is difficult to distinguish from
the SM (again within $2\sigma$), unless the present SM calculation
($Br(B\to X_s\gamma)= (1.9\pm 0.6)\times 10^{-4}$ \cite{ciuetal2}) is
improved \cite{bsg_th}.

\begin{figure}
\centering
\epsfxsize=4.5in
\leavevmode\epsffile{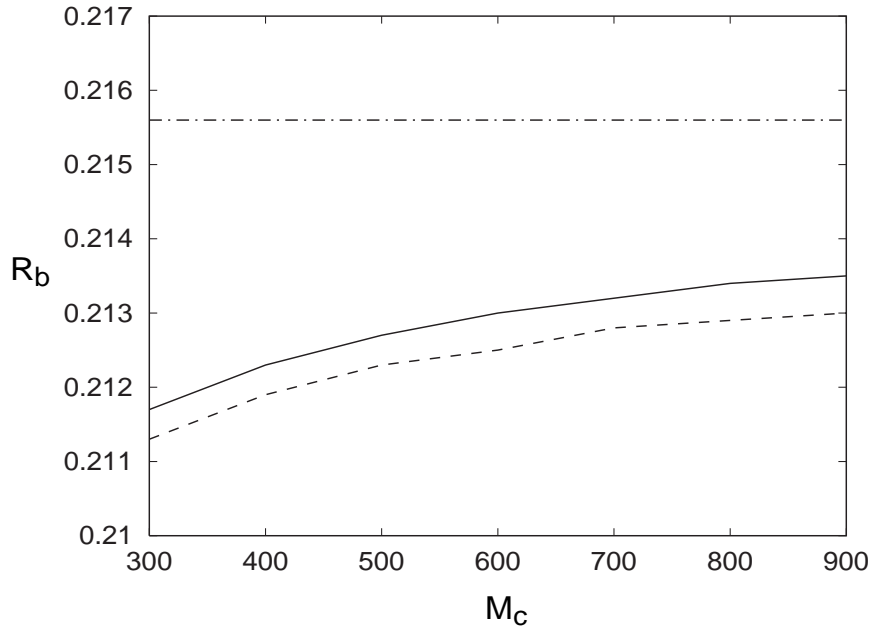}
\caption[]{$R_b$ in Model III for $\alpha=0$ (solid) and
$\alpha=\pi/4$ (dashed). The SM prediction $R_b^{\sss{\rm SM}}=0.2156$ 
is also given (dot-dashed) for comparison.}
\label{fig_rb}
\end{figure}

\noindent With the requirement of a large $M_c$ coming from $Br(B\to
X_s\gamma)$, we need to consider the discussion of $\rho_0$ again and
modify it accordingly. The charged scalar cannot be the heaviest
scalar particle anymore, otherwise $\Delta\rho_0^{\sss{\rm NEW}}$
would be as in eq. (\ref{deltarho}) and would contradict the present
global fit result (see eq. (\ref{rhozero_gf})). As already noted in
ref. \cite{grant} for Model II, there are two other possible scenarios

\be
M_H,M_h\le M_c\le M_A\,\,\,\,\,\mbox{and}\,\,\,\,
M_A\le M_c\le M_H,M_h
\label{Mc_inbetween}
\ee

\noindent in which $\Delta\rho_0^{\sss{\rm NEW}}$, as given by
eq. (\ref{deltarhozero}), turns out to be negative, and has in this
way the extra advantage of cancelling the effect of the top quark SM
contribution (see eq. (\ref{rhotop})). We note that none of the
previous scenarios would be compatible with an enhanced value of
$R_b$, because in that case $M_A$ and $M_h$ would be required to be
equal and light (see Section \ref{rb}).
\begin{figure}
\centering
\epsfxsize=4.5in
\leavevmode\epsffile{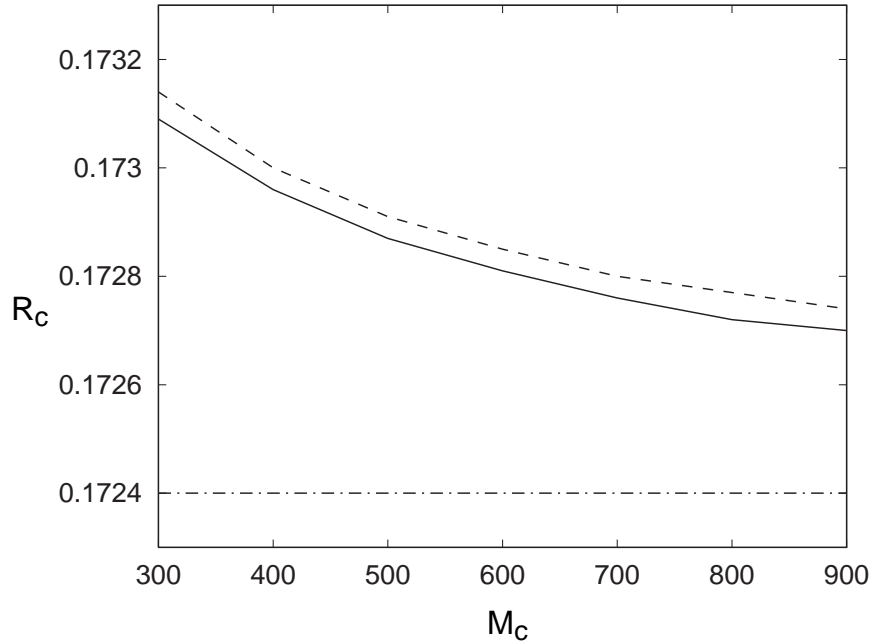}
\caption[]{$R_c$ in Model III for $\alpha=0$ (solid) and
$\alpha=\pi/4$ (dashed). The SM prediction $R_c^{\sss{\rm SM}}=0.1724$ 
is also given (dot-dashed) for comparison.}
\label{fig_rc}
\end{figure}

\begin{figure}
\centering
\epsfxsize=4.2in
\leavevmode\epsffile{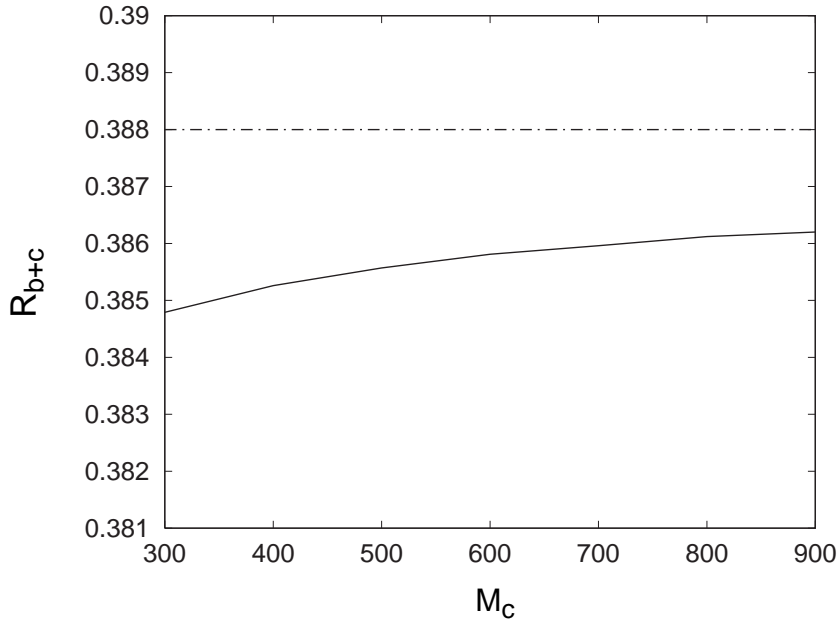}
\caption[]{$R_{b+c}$ in Model III (solid) compared to the SM
prediction (dot-dashed). The dependence on $\alpha$ is irrelevant.}
\label{fig_rbrc}
\end{figure}

{}From a direct numerical evaluation of $\Delta\rho_0^{\sss{\rm
NEW}}$, we find that there may exist many possible sets of mass
parameters for which eq. (\ref{rhozero_gf}) can be satisfied. For
instance, let us consider the case in which $M_H,M_h\le M_c\le
M_A$. The other case in eq. (\ref{Mc_inbetween}) has been studied too
and it gives comparable results. In order to have a small
$\Delta\rho_0^{\sss{\rm NEW}}$, it is crucial that $M_c$ and $M_A$ are
not too far apart. One possible optimal set of values for the mass
parameters is given by the following ratios: $M_H\!=\!0.4\, M_c$,
$M_h\!=\!0.5\, M_c$ and $M_A\!=\!1.1\, M_c$. In this case, the results
for $R_b$, $R_c$ and $R_{b+c}$ are illustrated in Fig. \ref{fig_rb} --
Fig. \ref{fig_rbrc} respectively. The SM predictions are also plotted
for comparison.  Clearly, in Model III, $R_b$ is less than
$R_b^{\sss{\rm SM}}$ and $R_c$ is larger than $R_c^{\sss{\rm
SM}}$. Thus, if the current experimental trend for $R_b^{\rm exp}$
exceeding $R_b^{\sss{\rm SM}}$ persists, Model III will be ruled out.

\section{Conclusions}
\label{concl}

We analyzed the decays $Z\rightarrow b\bar b$ and $Z\rightarrow c\bar
c$ in 2HDM with FCSC, often called Model III. We find that $R_b^{\rm
exp}$ places severe constraints on this model. It requires that
$M_h\sim M_A\le 60$ GeV, with significantly enhanced coupling of the
neutral scalar and pseudoscalar to $b\bar b$. This parameter space of
the model cannot be reconciled with constraints from the
$\rho$-parameter and $Br(B\rightarrow X_s\gamma)$.

Since aspects of the experimental analysis are of some concern, we
also examined the model by disregarding $R_b^{\rm exp}$ and we give
the predictions for $R_b$, $R_c$ and $R_{b+c}$ in this case. In
particular, we find that, if the current trend of $R_b^{\rm exp}>
R_b^{\sss{\rm SM}}$ persists, then this class of models will be ruled
out.

We emphasized the importance of $R_{b+c}$ and $R_l$ in our
analysis.

In view of the fact that in models with FCSC the rate for
$Z\rightarrow c\bar c$ receives a correction which grows with $m_t^2$,
we stress that precise measurements of $Z\rightarrow c\bar c$ could
provide unique constraints on the crucial $tc$-vertex.

\section*{Acknowledgments}

We acknowledge useful conversations with Louis Lyons, Vivek Sharma and
Shlomit Tarem. This research was supported in part by U.S. Department
of Energy contracts DC-AC05-84ER40150 (CEBAF) and DE-AC-76CH0016
(BNL).

\newpage

\appendix
\section{Feynman rules for Model III}
\label{fr_mod3}

In this appendix we summarize the Feynman rules for Model III which
are used in many of the calculations presented in the paper.

\subsection{Fermion-Scalar couplings}
We present the Feynman rules for the couplings of the scalar fields
$H^1$ (neutral scalar), $H^2$ (neutral pseudoscalar) and $H^+$
(charged scalar), to up-type and down-type quarks, as can be derived
from the Yukawa Lagrangian of Model III 
(eqs. (\ref{lyukmod3})-(\ref{lyukfc})). Following the discussion of
Section \ref{model}, these are the Feynman rules we need in our
calculation of $R_b$.

\begin{tabular}{p{4cm} p{8.2cm}}\\ \\ \\
\parbox[b]{4cm}{\epsffile{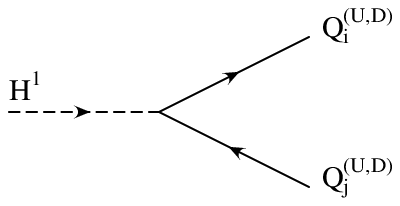}} & 
\raisebox{5.ex}{$\frac{-i}{2\sqrt{2}}\left((\xi_{ij}^{U,D}+
\xi_{ij}^{U,D*})+(\xi_{ij}^{U,D}-\xi_{ij}^{U,D*})\gamma_5\right)$}
\\ \\ \\
\parbox[b]{4cm}{\epsffile{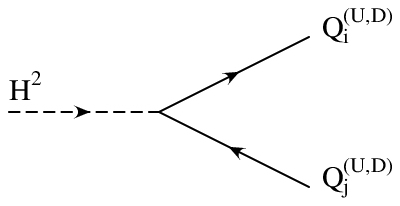}} & 
\raisebox{5.ex}{$\frac{1}{2\sqrt{2}}\left((\xi_{ij}^{U,D}-
\xi_{ij}^{U,D*})+(\xi_{ij}^{U,D}+\xi_{ij}^{U,D*})\gamma_5\right)$}
\\ \\ \\
\parbox[b]{4.cm}{\epsffile{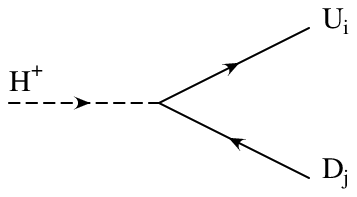}} & 
\raisebox{5.ex}{$\frac{-i}{2}\left(V_{\sss{\rm  CKM}}\!\cdot\!
\xi_{ij}^{D}(1+\gamma_5)-\xi_{ij}^{U}\!\cdot\! V_{\sss{\rm CKM}}
(1-\gamma_5)\right)$}\\ \\ \\
\end{tabular}

Although the $\xi_{ij}^{U,D}$ couplings are left complex in the above,
in practice, in our calculation we assumed they are real,
i.e. $\xi_{ij}^{U,D}\simeq\xi_{ij}^{U,D*}$, as we were not concerned
with any phase-dependent effects.

\subsection{Gauge boson-Scalar couplings}

Here is a list of the Z- and W-boson interactions with Model III
scalar fields, useful for the computation of $\Delta\rho_0^{\sss {\rm
NEW}}$.  We report them in terms of scalar mass eigenstates, $\bar
H^0$, $h^0$, $A^0$ and $H^+$, in order to make contact with the
discussion given in Section \ref{rho} and with the literature
\cite{bert,knowles}. We always have to remember the relations (see
eqs. (\ref{masseigen}) and (\ref{nomasseigen})) between the scalar
mass eigenstates and ($H^0$, $H^1$, $H^2$, $H^+$) and use the fact
that neither $Z H^0 H^1$ nor $Z H^0 H^2$ couplings are present
\cite{bert,knowles}. 

\begin{tabular}{p{6cm} p{5cm}}\\ \\ \\
\parbox[b]{6cm}{\epsffile{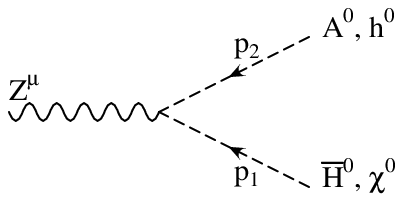}} & 
\raisebox{5.ex}{$\frac{g_{\sss W}}{2c_{\sss W}}\sin\alpha\,
(p_2-p_1)^{\mu}$}\\ \\ \\
\parbox[b]{6cm}{\epsffile{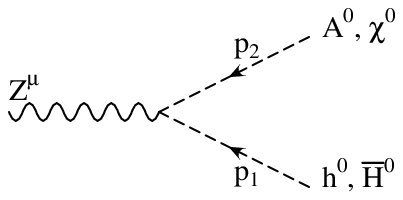}} & 
\raisebox{5.ex}{$\frac{g_{\sss W}}{2c_{\sss W}}\cos\alpha\,
(p_2-p_1)^{\mu}$}\\ \\ \\
\parbox[b]{6cm}{\epsffile{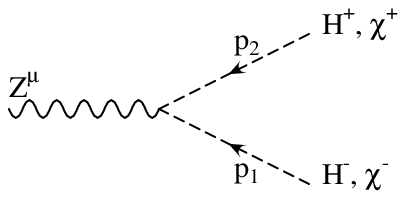}} & 
\raisebox{5.ex}{$\frac{ig_{\sss W}}{2c_{\sss W}}(1-2s_{\sss W}^2)
(p_2-p_1)^{\mu}$}\\ \\ \\
\parbox[b]{6cm}{\epsffile{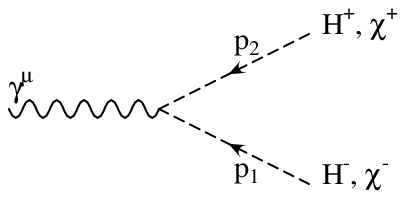}} & 
\raisebox{5.ex}{$ie\,(p_2-p_1)^{\mu}$}\\ \\ \\
\end{tabular}

\begin{tabular}{p{6cm} p{5cm}}\\ \\ \\
\parbox[b]{6cm}{\epsffile{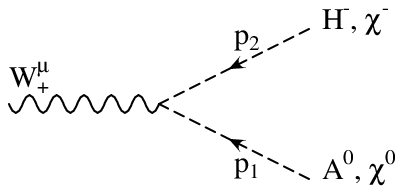}} & 
\raisebox{5.ex}{$\frac{g_{\sss W}}{2}(p_2-p_1)^{\mu}$}\\ \\ \\
\parbox[b]{6cm}{\epsffile{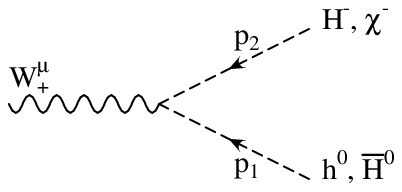}} & 
\raisebox{5.ex}{$\frac{-ig_{\sss W}}{2}\cos\alpha\,(p_2-p_1)^{\mu}$}\\ \\ \\
\parbox[b]{6cm}{\epsffile{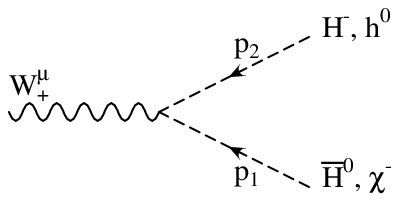}} & 
\raisebox{5.ex}{$\frac{-ig_{\sss W}}{2}\sin\alpha\,(p_2-p_1)^{\mu}$}\\ \\ \\
\end{tabular}

\newpage
\section{Calculation of $R_\ell$ as a function of $\alpha_s$}
\label{Rl}

In this Appendix we will use the value of $\alpha_s(M_Z)$ deduced from
physics other than the width for $Z\to{}$hadrons to predict
$\Gamma^{\sss{\rm SM}}(Z\to{}$hadrons) and $R^{\sss{\rm SM}}_\ell$ to
$0(\alpha^2_s)$. Mostly, we follow Bernab\'eu {\it et al}. \cite{pich},
who give expressions for various corrections to $\Gamma(Z\to f\bar
f)$, for both quarks and leptons.

Let us rewrite the expression for the width of $Z\to f\bar f$ as

\be
\Gamma(Z\to f\bar f)=
\Gamma^f_0(1+\Delta_{\rm EW}^f) (1+\Delta_{\rm QCD}^f)
\ee

\noindent where $\Gamma^f_0$ is the tree level expression, in which
some effects of the EW corrections have been reabsorbed in the
renormalization of the couplings (see conventions adopted in
\cite{pich}).  $\Delta_{\rm EW}^f$ includes only corrections which do
not depend on $\alpha_s$, i.e. pure EW corrections and QED
corrections. They are presented in detail in ref. \cite{pich}
(eqs. (9), (15) and (17), see also references therein) and we will not
discuss them here. We give their numerical values \cite{delta_ew} in
Table \ref{del_ew}.
\begin{table}
\begin{center}
\begin{tabular}{c c c c c}
\hline\hline
$\nu$ & $e,\mu,\tau$ & $u,c$ & $d,s$  & $b$\\ \hline
3.739 & 2.736 & 2.200 & 2.778 & -13.848  \\ \hline\hline
\end{tabular}
\caption[]{Values of $\Delta_{\rm EW}^f$, for different flavors, in
units of ($10^{-3}$). They have been evaluated for $m_t=176$ GeV and
$m_H=200$ GeV.}
\label{del_ew}
\end{center}
\end{table}
$\Delta_{\rm QCD}^f$ represents mostly $\alpha_s$-dependent
corrections which can be subdivided as

\be
\Delta^f_{\rm QCD} = \delta_{\rm QCD} + \delta^f_{\mu\rm QCD} + 
\delta^f_{t\rm QCD}
\ee

\noindent We briefly discuss each of them below.

The strong corrections to the basic $V,A$ vertex ($V=\gamma^\mu$,
$A=\gamma^\mu\gamma^5$) are flavor-independent and at $O(\alpha_s^2)$
are given by

\be
\delta_{\rm QCD} = \frac{\alpha_s(M_Z)}{\pi} + 1.41
\left(\frac{\alpha_s(M_Z)}{\pi}\right)^2 
\ee

\noindent This is the dominant effect amounting to about 3--4\% (see
Table \ref{delta_QCD}). 

$\delta^f_{\mu\rm{QCD}}$ represents corrections due to kinematic
effects of external masses, including mass-dependent QCD corrections
\cite{chet,kniehl}. We decide to include in the same factor also non QCD
mass-dependent corrections to the axial vector couplings, in order to
make the presentation more compact. Strictly speaking, this correction
should be included in $\Delta^f_{\rm EW}$. Based on the results given
in ref. \cite{chet,kniehl}, we deduce \cite{deltamu}

\be
\delta^f_{\mu\rm{QCD}} = \frac{3\mu^2_f}{v^2_f+a^2_f} \left[
-\frac{1}{2} a^2_f\left( 1+\frac{11}{3}\, \frac{\alpha_s}{\pi} \right) 
+ v^2_f\left(\frac{\alpha_s}{\pi} \right) \right]
\label{deltamuQCD}
\ee

\noindent where $\mu^2_f = 4\bar m^2_f(M_Z)/m^2_Z$, $\bar m_f(M_Z)$
being the running mass at the Z-scale, and

\bea
v_e & = -1+4x_{\sss{\rm W}} \quad , \quad a_e & = +1 \nonumber \\
v_u & = +1-\frac{8}{3}x_{\sss{\rm W}} \quad , \quad a_u & = -1 \\
v_d & = -1 + \frac{4}{3}x_{\sss{\rm W}} \quad , \quad a_d & = +1 
\nonumber 
\eea

\noindent Using eq. (2) from ref. \cite{pich}, we obtain $x_{\sss{\rm
W}}=.2314$ (where $x_{\sss{\rm W}}=\sin^2\!\theta_{\rm W}$).
Numerically, $\delta^b_{\mu\rm{QCD}} \simeq -5\times 10^{-3}$ and
$\delta^c_{\mu\rm{QCD}} \sim -0.5\times 10^{-3}$ (see Table
\ref{delta_QCD} for their $\alpha_s$-dependence). This kind of
correction is also relevant, without $O(\alpha_s)$ terms, for the
$\tau$ lepton, in which case it amounts to $\delta^\tau_\mu
\simeq-2\times 10^{-3}$.

At $O(\alpha^2_s)$ the large mass splitting between the $t$ and $b$
quarks gives rise to a correction, $\delta^f_{t{\rm QCD}}$, due to
triangular quark loops affecting the axial vector current
\cite{kniehl}:

\be
\delta^f_{t\rm QCD} = -\frac{a_ta_f}{v^2_f +a^2_f} \left(
\frac{\alpha_s}{\pi} \right)^2 f(\mu_t)
\ee

\noindent where $f(\mu_t)$ can be written as \cite{kniehl,pich}

\be
f(\mu_t)=\log\frac{4}{\mu_t^2}-3.083+0.346\frac{1}{\mu_t^2}+
0.211\frac{1}{\mu_t^4}
\ee

\noindent For $m_t=176$ GeV we use $f(\mu_t)=-4.374$. Thus,
this correction effects $+2/3$ charge-quarks positively and $-1/3$
charge-quarks negatively and for each flavor it is about
0.4-0.5\%, as we can read from Table \ref{delta_QCD}.
\begin{table}
\begin{center}
\begin{tabular}{|c|c|c|c|c|c|}
\hline\hline
$\alpha_s(M_Z)$ & $\delta_{QCD}$ & $\delta^b_{\mu\rm{QCD}}$ & 
$\delta^c_{\mu\rm{QCD}}$ & $\delta^u_{t\rm{QCD}}$ &
$\delta^d_{t\rm{QCD}}$  \\ \hline
0.105 & 34.998 & -5.417 & -0.560 & 4.260 & -3.305 \\ 
0.110 & 36.742 & -5.179 & -0.514 & 4.676 & -3.628 \\
0.115 & 38.495 & -4.938 & -0.467 & 5.111 & -3.965 \\
0.120 & 40.254 & -4.695 & -0.420 & 5.565 & -4.317 \\
0.125 & 42.021 & -4.450 & -0.372 & 6.038 & -4.684 \\ 
\hline\hline
\end{tabular}
\caption[]{Values of different QCD corrections (in units of
$10^{-3}$), for different values of $\alpha_s(M_Z)$.}
\label{delta_QCD}
\end{center}
\end{table}

Having identified all the corrections to $\Gamma_f\!=\!\Gamma(Z\to
f\bar f)$, for both quarks and leptons, we then consider $R_\ell$ and
define 

\bea 
R_\ell &=& \frac{(\Gamma_u + \Gamma_d + \Gamma_s +\Gamma_c
+\Gamma_b)} {\Gamma_\ell} \\ 
&=& \sum_{f=u,d,s,c,b}R^f_{\ell,0}\frac{(1+\Delta^f_{\rm EW})}
{(1+\Delta^\ell_{\rm EW}+\delta^{\tau}_{\mu}/3)}(1+\Delta^f_{\rm QCD})
\nonumber
\eea

\noindent where $\Gamma_\ell =(\Gamma_e+\Gamma_\mu+\Gamma_\tau)/3$ and
$\Delta^\ell_{\rm EW}$ represents the EW corrections common to all the
lepton species (see Table \ref{del_ew}). We have denoted by
$R^f_{\ell,0}$ the tree level ratios for each quark species. They are
given by

\bea 
R^u_{\ell,0} &=& \frac{\Gamma^u_0}{\Gamma^e_0} = 
3 \frac{v^2_u + a^2_u}{v^2_e + a^2_e} \nonumber \\
R^d_{\ell,0} &=& \frac{\Gamma^d_0}{\Gamma^e_0} = 
3 \frac{v^d_d + a^2_d}{v^2_e + a^2_e} 
\eea

\noindent and for $x_{\sss{\rm W}}=.2348$ they can be estimated to be
$R^u_{\ell,0}=3.4209$ and $R^d_{\ell,0}=4.4101$.

\noindent Finally, $\Delta^f_{\rm QCD}$ represents the total QCD
corrections for each flavor. They are deduced from the previous
discussion and their numerical values are summarized in Table
\ref{Delta_QCD}, together with $R_\ell$, for different values of
$\alpha_s(M_Z)$.

Using the values for $\Delta^f_{\rm EW}$ given, for each flavor, in
Table \ref{del_ew}, $R_\ell$ can be parametrized as follows
($\alpha_s=\alpha_s(M_Z)$) 

\bea
R_\ell &=& R^u_{\ell,0}\,(1.000219)\,(2+\Delta^u_{\rm QCD}(\alpha_s)+
\Delta^c_{\rm QCD}(\alpha_s))+ \nonumber \\
& & 2 R^d_{\ell,0}\,(1.000796)\,(1+\Delta^{d,s}_{\rm QCD}(\alpha_s)) + \\ 
& & R^d_{\ell,0}\,(0.984199)\,(1+\Delta^b_{\rm QCD}(\alpha_s)) \nonumber
\eea

\noindent from where we deduce the values reported in Table
\ref{Delta_QCD}.

\begin{table}
\begin{center}
\begin{tabular}{|c|c|c|c|c|c|}
\hline\hline 
$\alpha_s(M_Z)$ & $\Delta^u_{\rm QCD}$ & $\Delta^c_{\rm QCD}$ & 
$\Delta^{d,s}_{\rm QCD}$ & $\Delta^b_{\rm QCD}$ & $R_\ell$ \\ \hline \hline
0.105 & 39.258 & 38.698 & 31.693 & 26.276 & 20.6715 \\
0.110 & 41.418 & 40.904 & 33.114 & 27.935 & 20.7060 \\
0.115 & 43.606 & 43.139 & 34.530 & 29.592 & 20.7410 \\
0.120 & 45.819 & 45.399 & 35.937 & 31.242 & 20.7759 \\
0.125 & 48.059 & 47.678 & 37.337 & 32.887 & 20.8108 \\ \hline\hline
\end{tabular}
\caption[]{Values of $R_l$ and its QCD corrections (in units of
$10^{-3}$) as functions of $\alpha_s(M_Z)$.}
\label{Delta_QCD}
\end{center}
\end{table}


\begin{thebibliography}{99}
\bibitem{firstfamily} A detailed phenomenological analysis \cite{long}
shows that for the first family one needs $\lambda_{ij}\simeq 0.1$ in
eq. (\ref{coupl_sher}) in order to satisfy the bounds from $K^0$-$\bar
K^0$ and $B^0$-$\bar B^0$ mixing over a large region of the parameter
space of Model III. Therefore, in this work we will assume that the FC
couplings which involves the first family are negligibly small.
\bibitem{massless} In our calculation all the quarks except the top
quark are taken to be massless, keeping $m_q\ne0$ only in the
$\xi^{U,D}_{ij}$ couplings as in eq. (\ref{coupl_sher}) and in the
determination of some corrections to $R_\ell$, when they become
relevant to the degree of accuracy that we seek.
\bibitem{zbbzcc} In obtaining the result in eq. (\ref{rb_mod3})
we had calculated both $Z\to b\bar b$ and $Z\to c\bar c$ in Model III.
Therefore the corresponding experimental number differs a little bit
from that in eq. (\ref{rbexp}) which is determined by holding $R_c$
fixed to its experimental value (see ref. \cite{eps}).
\bibitem{alpha0} Note that we are considering the case $\alpha=0$, 
as required by the best fit of $R_b$.
\bibitem{normal} We use the same normalization as in
ref. \cite{buras}, in order to check our result against the discussion
of the problem in Model II, as given in that paper.
\bibitem{bsg_th} The theoretical prediction of $Br(B\rightarrow
X_s\gamma)$ from ref. \cite{ciuetal2} includes some NLO QCD
corrections and the result could change in the future by a complete
NLO analysis. We could have used in our analysis the fully consistent
LO result for $Br(B\rightarrow X_s\gamma)$, which is a little higher
(see ref. \cite{ciuetal2,buras}), but this would not modify the
qualitative results that we are giving.
\bibitem{delta_ew} $\Delta_{\sss EW}^f$ depend very weakly on
$m_H$. The numerical values quoted in Table  \ref{del_ew} are for
$m_H=200$ GeV.
\bibitem{deltamu} Note that our eq. (\ref{deltamuQCD}) differs from
eq. (19) of ref. \cite{pich}. We derive eq. (\ref{deltamuQCD}) from
ref. \cite{chet}, where the results of ref. \cite{kniehl} are improved
by absorbing large $\log{M_Z^2/m_b^2}$ through the use of the running
mass $\bar m_b(M_Z)$. Using eqs. (2) and (4) of ref. \cite{chet} and
substituting the on-shell mass $m_b$ with the $\overline{MS}$ mass as
in eq. (9) therein, we confirm their result in eq. (14). Thus we
arrive to our eq. (\ref{deltamuQCD}).
\end{thebibliography}
\end{document}